\documentclass[conference]{IEEEtran}
\pdfoutput=1
\usepackage{cite}
\usepackage{amsmath,amssymb,amsfonts,amsthm}
\usepackage{algorithmic}
\usepackage[ruled]{algorithm2e}
\usepackage{graphicx}
\usepackage{textcomp}
\usepackage{xcolor}
\usepackage{bbm}
\usepackage{url}
\usepackage{subcaption}
\usepackage[font=small,labelfont=bf]{caption}
\def\BibTeX{{\rm B\kern-.05em{\sc i\kern-.025em b}\kern-.08em
    T\kern-.1667em\lower.7ex\hbox{E}\kern-.125emX}}
    
\SetKwIF{If}{ElseIf}{Else}{if}{}{else if}{else}{end if}
\SetKwFor{While}{while}{}{end while}
\SetKwFor{For}{for}{}{}

\newcommand{\AC}{\mathcal{A}}

\newcommand{\CC}{\mathcal{C}}
\newcommand{\DC}{\mathcal{D}}
\newcommand{\EC}{\mathcal{E}}
\newcommand{\FC}{\mathcal{F}}
\newcommand{\GC}{\mathcal{G}}

\newcommand{\KC}{\mathcal{K}}
\newcommand{\LC}{\mathcal{L}}

\newcommand{\PC}{\mathcal{P}}

\newcommand{\XC}{\mathcal{X}}

\newcommand{\aB}{\mathbf{a}}

\newcommand{\cB}{\mathbf{c}}

\newcommand{\xB}{\mathbf{x}}
\newcommand{\yB}{\mathbf{y}}

\newcommand{\XB}{\mathbf{X}}

\DeclareMathOperator*{\argmin}{arg\,min}
\DeclareMathOperator*{\arginf}{arg\,inf}

\newtheorem{theorem}{Theorem}
\newtheorem{lemma}{Lemma}
\newtheorem{definition}{Definition}

\newcommand{\EntCond}[1]{H_c\left(#1 \right)}

\newcommand{\DivCond}[2]{D_c\left( #1 \| #2 \right)}

\newcommand{\LR}[3]{\left #1 {#2} \right #3}

\renewcommand{\Pr}{{\rm Pr}}

\newcommand{\px}{{\hat p_{\mathbf x}}}
 
\graphicspath{{../figs}}

\begin{document}

\title{The Metagenomic Binning Problem: \\ Clustering Markov Sequences}

\author{\IEEEauthorblockN{Grant Greenberg}
\IEEEauthorblockA{
Electrical and Computer Engineering Department \\
\textit{University of Illinois at Urbana-Champaign}\\
Champaign, IL, USA \\
gcgreen2@illinois.edu\vspace{-5mm}}
\and
\IEEEauthorblockN{Ilan Shomorony}
\IEEEauthorblockA{
Electrical and Computer Engineering Department \\
\textit{University of Illinois at Urbana-Champaign}\\
Champaign, IL, USA \\
ilans@illinois.edu}
}

\maketitle

\begin{abstract}
The goal of metagenomics is to study the composition of microbial communities, typically using high-throughput shotgun sequencing.
In the metagenomic binning problem, we observe random substrings (called contigs) from a mixture of genomes and want to cluster them according to their genome of origin.
Based on the empirical observation that genomes of different bacterial species can be distinguished based on their \emph{tetranucleotide frequencies}, we model this task as the problem of clustering $N$ sequences generated by $M$ distinct Markov processes, where $M \ll N$.
Utilizing the large-deviation principle for Markov processes, we establish the information-theoretic limit for perfect binning.
Specifically, we show that the length of the contigs must scale with the inverse of the Chernoff Information between the two most similar species. 
Our result also implies that contigs should be binned using the conditional relative entropy as a measure of distance, as opposed to the Euclidean distance often used in practice.
\end{abstract}



\section{Introduction}
    

In the last decade, advances in high-throughput DNA sequencing technologies have allowed a vast amount of genomic data to be generated.
Countless tasks such as genome assembly, RNA quantification, and genome-wide association studies have become a reality,
opening up exciting new research directions within biology and medicine \cite{Land2015}.

Significant attention has recently been given to the analysis of the human microbiome through \emph{metagenomics} \cite{Breitwieser}.
In metagenomics, a sample is taken from a microbial community, such as the human gut. 
The genetic material in the sample is then sequenced and analyzed to determine the microbial composition of the community \cite{Pachter_Chen}.
Recent research, including the Human Microbiome Project \cite{Chatelier}, has shown 
that the composition of the microbiome is a ``snapshot'' into an individual's overall health, providing 
great potential for personalized medicine.

Full reconstruction of the genomes in a metagenomic sample is generally infeasible due to insufficient coverage and high similarity across species \cite{MetaBAT}. 
In the typical analysis pipeline, the millions of reads obtained via high-throughput sequencing are used to create a much smaller number of contiguous sequences, known as \textit{contigs}, by merging reads with large overlaps \cite{Nissen}. 
The set of resulting contigs typically make up only a small fraction of the full genomes 
of all species present in the sample and have no significant overlaps with each other.




\par

Metagenomic binning is concerned with the following question:
is it possible to group the resulting contigs based on the genome from which they were derived?
Somewhat surprisingly, it has been shown that contigs belonging to the same species typically have similar sequence compositions.
Specfically, it was empirically verified that 
the distribution of four-letter strings (e.g. $AGCG$) remains relatively constant across an entire bacterial genome \cite{Noble,Mrazek}.
Hence one can compute for each contig the \emph{tetranucleotide frequency} (TNF) vector, and group together contigs with ``similar'' TNF vectors. 
Provided that the underlying TNF distributions are distinct enough, metagenomic binning can thus be performed.
Based on this idea\footnote{Usually, in addition to the TNF vector, the read coverage of each contig is used as another feature to help with the clustering. 
However, in this paper, we focus on using the TNF alone.}, many different algorithms and software packages have been proposed to perform metagenomic binning \cite{MetaBAT,Nissen,COCACOLA}. 
Other algorithms use a supervised learning approach by comparing the sequence composition of reads to a database of known bacterial genomes \cite{Kraken, OPAL, CLARK, metakallisto}, or through direct alignment to said database \cite{PhymmBL,MEGAN}.

The fact that the distribution of four-symbol strings is consistent throughout a given genome motivates the modeling of each genome as a \emph{third}-order Markov process.
Hence we assume that a contig is generated by one out of $M$ distinct, \emph{unknown} Markov processes $p_1,\dots,p_M$ with equal probability, where each $p_k$ corresponds to a certain species.
In order to study the fundamental limits of this problem, we assume all $N$ contigs have length $L$,
and consider an asymptotic regime where $N\to\infty$ and the contig length grows slowly with the number of contigs (specifically we set $L=\bar L \log N$).
Our goal is to characterize how large $\bar L$ needs to be in order to allow perfect binning with high probability.

To obtain our main result, we establish the equivalent of the Chernoff Information  \cite[Chapter~11.9]{Cover} for Markov processes, which gives the error exponent for the Bayesian error probability when testing between two known Markov processes.
This result, combined with a scheme to estimate the $M$ Markov distributions, allows us to show that perfect binning is possible if and only if
$$
\bar L>\frac1{\min_{k,\ell} C(p_k,p_\ell)},
$$
where $C(p_k,p_\ell)$ is the Chernoff Information between $p_k$ and $p_\ell$.
To estimate the unknown distributions, we consider building a graph where contigs whose empirical distributions are close are connected.
We then show that, with high probability, $M$ large cliques can be found, 
which can be used to find estimates $\tilde p_k$, $k=1,...,M$, of the Markov distributions.
Each contig $\xB$ is then then placed in bin $k$ given by
$$
\argmin_{k\in\{1,\dots,M\}} \DivCond{\hat p_\xB}{\tilde p_k}
$$
where $\hat p_\xB$ is the empirical 4-symbol distribution of $\xB$ and $D_c(\cdot \| \cdot)$ is the conditional relative entropy \cite[Definition 2.65]{Cover}.


Our main result suggests that the optimal way to bin metagenomic contigs is to estimate the underlying TNF distributions and then bin contigs using the conditional relative entropy as a metric, as opposed to the commonly used  Euclidean distance.
By simulating contigs from real bacterial genomes, we show that this metric can lead to lower binning error probabilities.

The paper is organized as follows.
In Section~\ref{sec:problem_statement} we describe the problem formulation in detail and state our main result.
In Section~\ref{sec:achievability} we describe our achievability scheme and the main technical ingredients used to prove it, and in Section~\ref{sec:converse} we describe the converse argument. 
In Section~\ref{sec:simulations} we provide preliminary simulation results, and we conclude the paper with a discussion in Section~\ref{sec:discussion}.

\section{Problem Statement}\label{sec:problem_statement}

As shown in \cite{Noble}, the distribution of tetranucleotides (four-letter strings), tends to be stationary across an individual bacterial genome.
Hence, it is natural to assume that each of the species in our sample corresponds to a distribution over all possible tetranucleotides\footnote{In practical approaches, \emph{reverse-complementary} tetranuclotides such as $ACAG$ and $CTGT$ are treated as the same tetranucleotide, but we ignore that fact for the sake of simplicity.
} 
$\{AAAA,AAAC,...,TTTT\}$.

Let ${\PC}$ be the $|\XC|^4$-dimensional simplex, where $\XC~=~\{A,C,G,T\}$.
Notice that not all distributions in $\mathcal P$ are valid tetranucleotide distributions, as the tetranucleotides in a sequence overlap with each other.
Let $\widetilde{\PC}$ be the set of all $p \in \PC$ with $p(\cB)>0,$  $\forall \, \cB \in \XC^4$, which, in addition, satisfy for all $\aB\in\XC^3$ 
\begin{equation}
\sum_{b \in \XC} p(\aB b) = \sum_{b \in \XC} p(b \aB). \label{eq:consistency}
\end{equation}
Condition (\ref{eq:consistency}) ensures that a given $p \in \widetilde{\PC}$ corresponds to the tetranucleotide distribution of a specific, \textit{stationary}, irreducible (due to $p(\cB)>0$), third-order Markov chain.
More precisely, we can let the induced distribution over $3$-letter strings be
\begin{equation}\label{eq:third_order_dist}
p(\aB) = \sum_{b \in \XC} p(\aB b).
\end{equation}
This uniquely determines a stationary Markov process with initial state distributed as (\ref{eq:third_order_dist}) and transition probabilities (i.e. conditional distribution) 
\begin{equation}\label{eq:conditional_dist}
p(b|\aB) = \frac{p(\aB b)}{p(\aB)}.
\end{equation}
Hence we will model each species in the sample using a distribution $p_k \in \widetilde{\PC}$.

%
%

%
%
%
%
%
%
%
%
%

\subsection{Metagenomic Binning Problem}
We assume that we have $M$ species in our sample (for a known $M$). 
Each species is modeled by a stationary third-order Markov process defined by $p_k \in \widetilde{\PC}$, for $k=1,...,M$.
From this genomic mixture, we observe a set of $N$ realizations $\CC=\{\xB_i\}_{i=1}^N$, which we call \textit{contigs}. 
Each $\xB\in\CC$ is generated independently by first choosing a species $k \in \{1,...,M\}$ with uniform prior probabilities $\frac1M$, and then generating a length-$L$ sequence according to $p_k$. 
For each $k$, let $\CC_k$ be the set of contigs generated according to $p_k$.
We wish to reconstruct $\CC_k,\ k=1,\dots,M$, by determining which contigs originated from the same genome.
\par
We point out that in real metagenomic experiments, the \emph{coverage depth},
that is, the expected number of contigs containing a specific nucleotide from one of the $M$ genomes, is low \cite{metaSPAdes}.
Hence, contigs will have no overlap with high probability, allowing us to model them as independent realizations of the different Markov processes in the sample.

\subsection{Perfect Binning}

The goal of the metagenomic binning problem is to cluster the $N$ contigs into $M$ ``bins", where each bin $k$ corresponds to a unique species with distribution $p_k$. 
More precisely, the goal is to find a decision rule $\delta: \XC^L\rightarrow \{1,\dots,M\}$ (using notation from \cite{Moulin}) which correctly maps each contig to its respective genome bin. 
\par
Perfect binning would be achieved if for every contig $\xB$, $\delta(\xB)$ chooses the label of the distribution from which it was generated.
However, we have the added difficulty that the distributions are unknown.
As a result, we can only require the decision rule to be correct up to a consistent relabeling of species indices. 
Hence the error event for a decision rule $\delta$ is 
\begin{equation}
    \EC_{\delta}=\{\exists \xB\in\CC_k,\yB\in\CC_\ell,k\ne\ell:\delta(\xB)=\delta(\yB)\}.
\end{equation}
We would like to know under what circumstances we can perfectly bin all $N$ contigs. 
In order to study the information-theoretic limits of this problem, we analyze an asymptotic regime, similar to \cite{Motahari}, in which $N\to\infty$ and
\begin{equation}\label{eq:scaling}
    L=\bar L \log N
\end{equation}
where $\bar L$ is the ``normalized contig length". 
Intuitively, a larger value of $\bar L$ should allow one to bin a contig with higher accuracy.
This scaling forces the contig length to be small compared to the number of contigs and, as we will show, is a meaningful scaling for the asymptotic problem we consider. 
 
This asymptotic regime allows us to define when species are resolvable as follows:
\par

\begin{definition}
    The $M$ species with distributions $\{p_k\}_{k=1}^M$ are \textbf{resolvable} if there exists a sequence of decision rules $\{\delta_N\}$ such that Pr$(\EC_{\delta_N})\rightarrow 0$ as $N\rightarrow\infty$.
\end{definition}

\subsection{Main Result}

Interestingly, the fundamental limit of resolvability relies on the Chernoff Information, which we define next.
\begin{definition}\label{def:chernoff}
For two Markov processes $p_k$ and $p_\ell$, the \textbf{Chernoff Information} between $p_k$ and $p_\ell$ is given by
    \begin{equation}\label{chernoff_info}
        C(p_k,p_\ell)=\DivCond{p^*}{p_k}=\DivCond{p^*}{p_\ell}
    \end{equation}
    where $D_c$ is the conditional relative entropy \cite[Def. 2.65]{Cover}, and $p^*$ is the solution to the following minimization problem.
    \begin{gather}
        p^*=\argmin_{p\in\widetilde{\PC}}\DivCond{p}{p_k} \nonumber \\ 
         \quad\text{s.t. } \DivCond{p}{p_k}=\DivCond{p}{p_\ell} \label{eq:pstar_chern}
    \end{gather}
\end{definition}

The Chernoff Information can be thought of as a measure of distance between the distributions.
Our main result establishes that the minimum normalized contig length, $\bar L$, required for \textit{resolvability} depends exclusively on the minimum Chernoff Information between species distributions.
\begin{theorem}\label{thm:resolvable}
    Let $C_{\min}= \min_{k\ne\ell} C(p_k,p_\ell)$. The species' distributions $\{p_k\}_{k=1}^M$ are resolvable if and only if
    \begin{equation} \label{eq:thm}
        \bar L>\frac1{ C_{\min}}
    \end{equation}
\end{theorem}

Intuitively, this means that the contig length must be large enough to distinguish between the two closest distributions.

\section{Achievability}\label{sec:achievability}
The achievability proof of Theorem~\ref{thm:resolvable} is described in the form of an algorithm so as to highlight the algorithmic nature of metagenomic binning.
Given a contig $\xB\in\CC$, we define the empirical fourth-order distribution of $\xB$ as $\hat p_\xB$ and we use $d$ as the $\ell_1$ 
distance between distributions, i.e. $d(p,q)~=~\sum_{\cB\in\XC^4}~|p(\cB)-q(\cB)|$. 
\begin{algorithm}
\SetAlgoLined
\SetAlgoNoEnd
\DontPrintSemicolon
\KwResult{Decision Rule $\delta(\xB)$}
\KwIn{Contigs $\CC$, Parameter $\alpha\in(0,1)$}
\Begin{ 
    $\DC \longleftarrow \text{sort in ascending order}\LR\{{d(\hat p_{\xB},\hat p_{\yB}),\forall\xB,\yB\in\CC}\}$\;
    \For {$\epsilon \text{ in } \DC$}{
        $\GC_\epsilon\longleftarrow\LR({V=\CC,E_\epsilon=\{(\xB,\yB): d(\hat p_\xB, \hat p_\yB)\leq\epsilon\}})$\;
        \If {$\GC_\epsilon$ \text{\normalfont has cliques} $\{\KC_k\}_{k=1}^M,|\KC_k|\geq (1-\alpha)\frac NM$} {
            \For{$k\longleftarrow 1$ \text{\normalfont to} $M$}{
                $\tilde  p_k\longleftarrow\frac1{|\KC_k|}\sum_{\xB\in \KC_k}\hat p_{\xB}$\;
            }
            \textbf{break}
        }
    }
    \For{$\xB\in\CC$}{  
        $\delta(\xB)\longleftarrow\argmin_{k\in\{1,\dots,M\}} \DivCond{\hat p_\xB}{\tilde p_k}$\;
    }
}
\caption{Binning Contigs}
\label{alg:achievability}
\end{algorithm}




The algorithm first estimates the species distributions by averaging the empirical distributions of the contigs in each clique, then it bins the contigs based on the estimates. Note that the algorithm as described is not computationally efficient (specifically, finding large cliques), and is used only to establish the achievability of Theorem \ref{thm:resolvable}. 

\subsection{Estimating Distributions}\label{subsection_estimate_distributions}
Recall that $\CC_k$ is the set of contigs generated by $p_k$.
We expect the empirical distribution of the majority of contigs in $\CC_k$ to be near $p_k$.
To identify those ``good'' contigs, let 
$$ \CC_{k,\epsilon}  = \{ \xB \in \CC_k : d(\hat p_{\xB},p_k) \leq \epsilon \}.
$$
To prove that the distribution estimates $\{\tilde p_{k}\}_{k=1}^M$ are close to the true distributions $\{p_k\}_{k=1}^M$ (after proper reindexing), we will first show that each clique the algorithm identifies is ``pure'' in the sense that it contains ``good" contigs from only a single species.
We will let  $d_{\min}~\triangleq~\min_{k\neq \ell}~d(p_k,p_\ell)$ be the minimum $\ell_1$ distance between any pair of the $M$ species distributions.

\begin{lemma}\label{lemma_pure_cliques}
If $\KC_k$ is a clique in $\GC_\epsilon$ for $\epsilon<  \frac{d_{\min}}2$, 
then
\begin{equation} 
\sum_{\ell=1}^M \mathbbm{1}\{\KC_k \cap \CC_{\ell,\epsilon/2} \ne \emptyset \} \leq 1,
\end{equation}
\end{lemma}

Lemma~\ref{lemma_pure_cliques} establishes that, if Algorithm~\ref{alg:achievability} finds $M$ cliques of size $(1~-~\alpha)\frac NM$ in $\GC_\epsilon$ for $\epsilon~<~\frac{d_{\min}}2$, then each clique contains contigs from at most one $\CC_{\ell,\epsilon/2},\ \ell=1,\dots,M$.
In order to establish that these $M$ cliques will exist in $\GC_{\epsilon}$, we use the following lemma, which essentially says that a large fraction of the contigs will be close to their respective generating distributions.

\par
\begin{lemma}\label{lemma_alphaNM_contigs}
For $\epsilon > 0$, $k \in \{1,...,M\}$, and $N$ large enough,
\begin{equation}\label{eq:lemma2}
\Pr\left( |\CC_{k,\epsilon/2}| < (1-\alpha)\frac NM \right) \leq \frac2\alpha e^{-\gamma\alpha^2 L},
\end{equation}
where $\gamma$ is a positive constant.
\end{lemma}


Fixing $\epsilon < \frac{d_{\min}}2$, Lemma~\ref{lemma_alphaNM_contigs} guarantees that for a reasonably chosen $\alpha$, as long as $N$ is large enough, we will have $ |\CC_{k,\epsilon/2}| \geq (1-\alpha)\frac NM$ for all $k~=~{1,...,M}$.
Moreover, by the triangle inequality, any two contigs $\xB,~\yB~\in~\CC_{k,\epsilon/2}$ will be at a distance $\epsilon$ or less of each other and will thus have an edge between them.
Hence, $\CC_{k,\epsilon/2}$ forms a clique in $\GC_\epsilon$.

Notice that $d_{\min}$ is not known, so the algorithm cannot restrict its search to $\epsilon < \frac{d_{\min}}2$.
However, since the algorithm considers different values of $\epsilon$ in increasing order, for some $\epsilon < \frac{d_{\min}}2$, $M$ cliques of size $(1-\alpha)\frac NM$ will exist with probability $1-o(1)$.
Lemma~1 will then guarantee that any cliques $\KC_1,...,\KC_M$ that are found will be pure.


\par
Consider a clique $\KC_k$ and let $\ell$ be such that $\KC_k\cap \CC_{\ell,\epsilon/2} \ne \emptyset$.
By Lemma~\ref{lemma_alphaNM_contigs}, the fraction of ``good" contigs in $\KC_k$ will be
\begin{equation}\label{eq:frac_of_good_contigs}
    \frac{|\KC_k\cap \CC_{\ell,\epsilon/2}|}{|\KC_k|}\geq 1-\frac{N-M\cdot(1-\alpha)\frac NM}{(1-\alpha)\frac NM}=1-\frac{\alpha M}{1-\alpha}
\end{equation}
with probability $1-o(1)$.
The lower bound results from dividing the maximum number of contigs \emph{not} in any clique by the minimum number of contigs in $\KC_k$.
If we set $\alpha=\frac1{\log L}$,  (\ref{eq:frac_of_good_contigs}) converges to $1$ and (\ref{eq:lemma2}) converges to $0$ as $N \to \infty$.
Thus, with high probability, a vanishing fraction of the contigs in $\KC_k$ does not belong to $\CC_{\ell,\epsilon/2}$.
Since distribution vectors are bounded, the impact of wrong contigs in $\KC_k$ on $\tilde p_k$ also vanishes, and we conclude that the distribution estimate 
$\tilde p_k = \frac1{|\KC_k|}\sum_{\xB\in \KC_k}\hat p_{\xB} \to p_\ell$ as $N \to \infty$.


\subsection{Binning Contigs}

In Subsection~\ref{subsection_estimate_distributions}, we established that we can construct estimates of the underlying distributions $\{p_k\}_{k=1}^M$ that are arbitrarily accurate as $N\to \infty$.
Next we show that, binning the contigs based on the conditional relative entropy using the underlying distributions achieves (\ref{eq:thm}) in the limit. 

Consider the hypothesis test between two Markov processes $p_k$ and $p_\ell$ (assumed to be known). Given prior probabilities $\pi_k$ and $\pi_\ell$, the Bayesian probability of error is
$$
    \pi_k\Pr(\text{choose }\ell|k\text{ true})+\pi_\ell\Pr(\text{choose }k|\ell\text{ true})
$$
for the decision rule on a contig generated by either $p_k$ or $p_\ell$.
\begin{theorem}\label{thm:chernoff}
    Let $\EC_{k,\ell}^{(L)}$ be the error event for the decision rule which minimizes the Bayesian probability of error. 
    Then
    \begin{equation}\label{eq:error_exponent_chernoff}
        \lim_{L\to\infty}\frac1L\log\Pr(\EC_{k,\ell}^{(L)}) = -C(p_k,p_\ell),
    \end{equation}
    i.e., $C(p_k,p_\ell)$ is the optimal error exponent.
\end{theorem}

The proof of Theorem \ref{thm:chernoff} is given in Section \ref{sec:proofs}. 
For a given contig, the last step of Algorithm \ref{alg:achievability} can be thought of as $M-1$ binary hypothesis tests between the true distribution and each of the remaining distributions. 
Thus, we will use Theorem \ref{thm:chernoff} to bound the overall error probability, $\Pr(\EC_{\delta_L})$, by considering the two closest distributions. 
\begin{align}
    \Pr(\EC_{\delta_L})&\leq\sum_{\xB\in\CC_k}\pi_k\sum_{k=1}^M\sum_{\ell \ne k}
        \Pr(\EC_{k,\ell}^{(L)}) \label{eq:union_ach}\\
    &\leq N\LR({\frac1M})M(M-1)\max_{k\ne\ell}\Pr(\EC_{k,\ell}^{(L)})\label{eq:n_contigs_ach}\\
    &\leq M 2^{L\LR({1/\bar L+\max_{k\ne\ell}(1/L)\log\Pr(\EC_{k,\ell}^{(L)})})} \label{eq:ach_end}
\end{align}
where (\ref{eq:union_ach}) follows from the union bound. 
By Theorem~\ref{thm:chernoff},  
\begin{align*}
    \max_{k\ne\ell}\frac1L\log\Pr(\EC_{k,\ell}^{(L)}) &\to -\min_{k\ne\ell}C(p_k,p_\ell) = -C_{\min}
\end{align*} 
as $N\to\infty$. 
Hence, if $\bar{L}>\frac{1}{C_{\min}}$,
$\Pr(\EC_{\delta_L}) \to 0$. 
Consider the case when instead we have estimates of the true distributions. 
The decision boundary for the optimal assignment of contigs is continuous on $\{\tilde p_k\}_{k=1}^M$. Since each $\tilde p_k\to p_\ell$, a continuity argument can be used to show that the probability of error of the binary hypothesis test converges to the same value as the distributions converge to the true ones. Then, it follows that the overall error probability converges to (\ref{eq:ach_end}).
This concludes the achievability proof of Theorem \ref{thm:resolvable}. 

\section{Converse}\label{sec:converse}
Without loss of generality, let $p_1$ and $p_2$ be such that $C_{\min}~=~C(p_1,p_2)$. Given the decision rule $\delta_L$, contigs $\xB_1\in\CC_1$ and $\xB_2\in\CC_2$, and a contig $\xB\in\CC$, let 
\begin{equation}
    \widetilde \EC_{1,2,\xB}=\{\xB\in\CC_1,\delta_N(\xB)\neq\delta_N(\xB_1)\}\cup\{\xB\in\CC_2,\delta_N(\xB)\neq\delta_N(\xB_2)\}
    \nonumber
\end{equation}
i.e. the event that $\xB$ was generated by either $p_1$ or $p_2$ \emph{and} incorrectly binned. 
Note that 
$\Pr(\widetilde \EC_{1,2,\xB})~\geq~\frac2M\Pr(\EC_{1,2})$. Then 
\begin{align}
    \Pr(\EC_{\delta_N}) & \geq \Pr\LR({\bigcup_{i=1}^N \widetilde \EC_{1,2,\xB_i}}) = 1-(1-\Pr(\widetilde\EC_{1,2,\xB_1}))^N \nonumber \\
    & \geq 1-\LR[{\LR({1-\frac2M\Pr(\EC_{1,2})})^{1/\Pr(\EC_{1,2})}}]^{N\Pr(\EC_{1,2})} \nonumber \\
    & \geq 1-e^{-\frac2MN\Pr(\EC_{1,2})} \label{eq:conv_poi_approx}
\end{align}
where (\ref{eq:conv_poi_approx}) follows from the bound $(1-ap)^{1/p}\leq e^{-a}$ for $p\in(0,1],a\in\mathbb R$. 
We see that, if $N\Pr(\EC_{1,2})~\not\to~0$, then $\Pr(\EC)~\not\to~0$. Since
$$
    N\Pr(\EC_{1,2})=2^{L\LR({1/{\bar L}+(1/L)\log\Pr(\EC_{1,2})})},
$$
then by Theorem~\ref{thm:chernoff}, $N\Pr(\EC_{1,2})\not\to0$ when $\bar L~\leq~\frac{1}{C_{\min}}$. 
This concludes the converse proof for 
Theorem~\ref{thm:resolvable}.

\section{Experimental Results}\label{sec:simulations}


From the point of view of practical metagenomic binning algorithms, our main result suggests that:
\begin{enumerate}
    \item the conditional relative entropy is a good metric for binning contigs,
    \item the Chernoff information can be used as a measure of how difficult it is to distinguish two species.
\end{enumerate}

In this section, we provide preliminary empirical evidence of these claims.
To this end, we utilized several previously sequenced and assembled bacterial genomes, available at NCBI \cite{NCBI}.
For each bacterial species $k$, we numerically computed its fourth-order distribution $p_k$ (i.e., the overall tetranucleotide frequency vector). We were able to simulate contigs of a desired length $L$ by sampling from all length-$L$ substrings from the genome.
For each experiment, we assume $N=10^6$ for concreteness (thus $L=\bar L \log 10^6$), but the results are not significantly affected by this choice.



\begin{figure*}
\vspace{-2mm}
    \begin{subfigure}{.33\textwidth}
        \centering
        \includegraphics[scale=0.3]{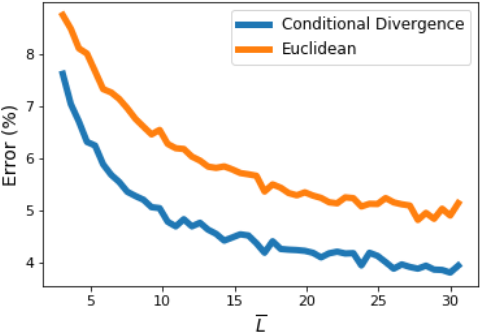}
        \caption{}
        \label{fig:div_vs_euc}
    \end{subfigure}
    \begin{subfigure}{.33\textwidth}
        \centering
        \vspace{-4mm}
        \includegraphics[scale=0.4]{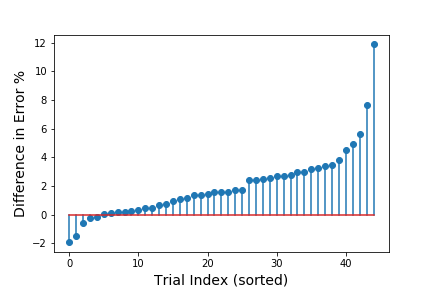}
        \caption{}
        \label{fig:all_pairs}
    \end{subfigure}
    \begin{subfigure}{.33\textwidth}
        \centering
        \includegraphics[scale=0.3]{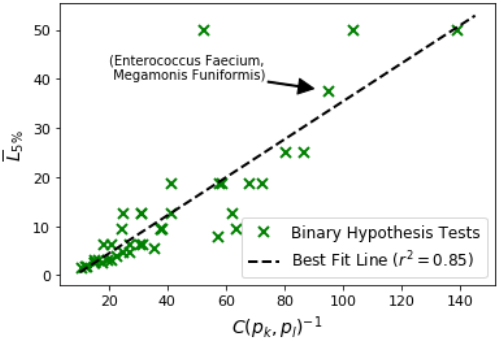}
        \caption{}
        \label{fig:scaling}
    \end{subfigure}
    \caption{\textbf{(a)} Comparison of conditional divergence and Euclidean distance for a hypothesis test between Alistipes Obesi and Megamonas Funiformis; \textbf{(b)} The difference between error percentages for Euclidean distance and conditional divergence with $\bar L=C(p_k,p_\ell)^{-1}$}; \textbf{(c)} Normalized contig length required for 5\% error ($\overline L_{5\%}$) vs the inverse of the Chernoff information for several pairs of species.
\end{figure*}

\par 
In order to verify the usefulness of the conditional relative entropy and compare it to the Euclidean distance (used in state-of-the-art tools such as \cite{MetaBAT,COCACOLA}), we considered the following experiment:
we extracted random contigs from a species $p_1$ and then tested whether it was closer to species $p_1$ or to another species $p_2$ based on both the Euclidean distance and the conditional relative entropy. 
In Figure \ref{fig:div_vs_euc}, the conditional relative entropy metric\footnote{$D_c(\cdot||\cdot)$  is not technically a metric as it is not symmetric.} consistently outperforms the Euclidean metric as we vary $\bar L$ in the test between the species Alistipes Obesi and Megamonas Funiformi.

We performed this experiment for 45 different choices of pairs of bacterial genomes from NCBI. For each pair ($k,\ell$), we considered a fixed normalized contig length given by $\bar L=C(p_k,p_\ell)^{-1}$.
As shown in Figure \ref{fig:all_pairs}, the conditional divergence improves the error compared to the Euclidean distance in almost $90\%$ of cases.


\par
Theorem~\ref{thm:resolvable} implies that the inverse of the Chernoff Information characterizes how long the extracted contigs need to be in order for two species to be reliably distinguishable.
In order to verify that, we calculated $\bar L_{5\%}$, the minimum normalized contig length required to guarantee a $5\%$ error rate in the Bayesian hypothesis test between species $p_k$ and $p_\ell$ with equal priors.
In Figure \ref{fig:scaling}, we plot
$\bar L_{5\%}$ vs $C^{-1}(p_k,p_\ell)$ for many such pairs and observe a roughly linear relationship between these two quantities.
Such a linear relationship agrees with the relationship suggested by Theorem 1. Moreover, it provides support to the claim that $C^{-1}(p_k,p_\ell)$ is a measure of how difficult it is to distinguish contigs from two species based on tetranucleotide frequencies.
\section{Discussion}\label{sec:discussion} 

In this paper, we modeled the metagenomic binning problem 
as the problem of clustering sequences generated by distinct Markov processes.
While overly simplistic, this model allowed us to establish the Chernoff Information as a measure of how easy it is to distinguish contigs generated by two species.

The algorithm used to prove the achievability suggests that a good ``metric'' for binning is the conditional relative entropy between a contig and an estimate of a species TNF.
Through experiments, we provided preliminary evidence that this metric often outperforms the Euclidean metric in the problem of assigning a contig to a species bin.
However, this assumes knowledge of the overall TNF of a genome, which is not known in practical settings. Therefore a natural direction for future investigation is how to efficiently estimate the TNF distribution for the species present in the sample.

Furthermore, it is unclear whether estimating the underlying TNF distributions is necessary to achieve the fundamental limit.
Alternatively, one could consider an approach that directly clusters the contigs based on their pairwise distances or based on a graph obtained by thresholding the distances (similar to our $\GC_\epsilon$).
We point out that, for such a graph, the problem becomes a community detection problem, and bears similarities with the stochastic block model \cite{abbe2015exact}, since for each species there is a given probability that an edge is placed among two of its contigs, and for each pair of distinct species, there is another probability that an edge is placed between their contigs. 
These probabilities would in general depend on the species TNF distributions (or the Markov processes generating the contigs). 
Notice that, unlike in the standard stochastic block model, here the placing of the edges would not be independent events.


Finally, we point out that in most approaches to metagenomic binning, the read coverage, or \textit{abundance}, is used to compare contigs in addition to the TNF. 
The read coverage of a contig is essentially the average number of reads that cover any given base in the contig.
Intuitively, this number is proportional to the abundance of the corresponding species in the mixture.
Hence, one expects contigs from the same species to have similar read coverages, which can be used to improve metagenomic binning.
Another direction for future work is thus to consider the metagenomic binning problem where the different species have different abundances and, for each contig, one observes a read coverage value that is related to the species abundance.


\par


\section*{Acknowledgements}

The idea of modeling the species in a metagenomic sample as distinct Markov processes and utilizing the large deviation principle to characterize their separability came from discussions with Xingyu Jin, Vasilis Ntranos, and David Tse.

\bibliographystyle{IEEEtran}
\bibliography{citations}

\section{Appendix}\label{sec:proofs}
\subsection{Proof of Lemma \ref{lemma_pure_cliques}}

Suppose by contradiction that $\xB, \yB \in \KC_j$, $\xB \in \CC_{k,\epsilon/2}$ and $\yB \in \CC_{\ell,\epsilon/2}$, for $k \ne \ell$.
Then $d(\xB,\yB) < \epsilon$, and we have
    \begin{align*}
        d(p_k,p_\ell)&\leq d(p_k,\hat p_{\xB})+d(\hat p_{\xB},p_\ell) \\
        &\leq d(p_k,\hat p_{\xB})+d(\hat p_{\xB},\hat p_{\yB})+d(\hat p_{\yB},p_l) \\
        & < \epsilon/2 + \epsilon + \epsilon/2 < d_{\min},
    \end{align*}
which is a contradiction to the definition of $d_{\min}$.

\subsection{Proof of Lemma \ref{lemma_alphaNM_contigs}}
Let $\EC_k$ be the event of interest, $\LR\{{|\CC_{k,\epsilon/2}| < (1-\alpha)\frac NM}\}$, and let $\AC_k=\LR\{{|C_k| < (1-\frac\alpha2)\frac NM}\}$. 
Note that we use $\frac\alpha2$ for $\AC_k$ as opposed to $\alpha$ because we need $|C_k|$ to be larger than $|C_{k,\epsilon/2}|$. 
By Hoeffding's inequality,
$$
    \Pr(\AC_k) \leq e^{-2\LR({\frac{\alpha}{2M}})^2N} = e^{-N\frac{\alpha^2}{2M^2}}
$$
This means, with high probability, $p_k$ will generate enough contigs.

\par

Let $\FC_k$ be the set of distributions ``far" from $p_k$:
\begin{equation}
    \FC_k=\LR\{{p\in\widetilde{\PC}:d\left(p_k,p\right)\geq \frac\epsilon2}\}.
\end{equation} 
By a version of Sanov's theorem for Markov chains, given in Theorem \ref{thm:sanovs}, for any $\xB~\in~\CC_k$,
\begin{equation}\label{sanovs_estimating}
    \Pr(\px\in \FC_k) \leq (L+1)^4 2^{-L\DivCond{p^*}{p_k}}
\end{equation}
where
$p^*=\arginf_{p\in \FC_k}\DivCond{p}{p_k}$; 
i.e., $p^*$ is the distribution in $\FC_k$ closest to $p_k$ in conditional relative entropy.
Notice that $\EC_k$ occurs when more than $|C_k|~-~(1~-~\alpha)\frac NM~+~1$ contigs lie in $\FC_k$, leaving an insufficient number of ``good" contigs. 
Letting $\xB_0\in\CC_k$ be some contig generated by $p_k$,
\begin{align}
    & \Pr(\EC_k|\AC_k^c) \nonumber \\
    = & \Pr\left(\sum_{\xB\in\CC_k}\mathbbm{1}\{\hat p_{\xB}\in \FC_k\}\geq |\CC_k|-(1-\alpha)\frac NM+1\middle|\AC_k^c\right) \nonumber \\
    \leq & \Pr\left(\sum_{\xB\in\CC_k} \mathbbm{1}\{\hat p_{\xB}\in \FC_k\}\geq \frac\alpha2\frac NM\middle|\AC_k^c\right) \label{eq:l2_sym}\\
    \leq & \frac{2M}{\alpha}\cdot\Pr(\hat p_{\xB_0}\in \FC_k|\AC_k^c) \label{eq:markov_ineq}
\end{align}
where (\ref{eq:l2_sym}) follows the definition of $\AC_k$, and (\ref{eq:markov_ineq}) from Markov's inequality and symmetry across contigs.
Combining the probabilities,
\begin{align}
    \Pr(\EC_k)&=\Pr(\EC_k|\AC_k^c)\Pr(\AC_k^c)+\Pr(\EC_k|\AC_k)\Pr(\AC_k) \\
    &\leq \frac{2M}{\alpha}\cdot\Pr(\hat p_{\xB_0}\in \FC_k|\AC_k^c)+\Pr(\AC_k) \\
    &\leq \frac{2M}{\alpha}(L+1)^4 2^{-L\DivCond{p^*}{p_k}} + e^{-N\frac{\alpha^2}{2M^2}} \label{eq:usegamma} \\
    &\leq \frac2\alpha e^{-\gamma \alpha^2 L} \label{eq:l2_simplification}
\end{align}
where $\gamma~>~0$ is a constant that does not depend on $\alpha$ or $L$, guaranteed to exist such that (\ref{eq:l2_simplification}) holds for large $N$. $\gamma$ can be found by manipulating (\ref{eq:usegamma}) using simple algebraic operations.

\subsection{Proof of Theorem \ref{thm:chernoff}}
We define the \textit{type} of a contig $\xB$ to be its empirical fourth-order distribution, denoted $\px$.
Let the set of all possible types of length-$L$, stationary, third-order Markov sequences be $\PC_L$. The cardinality of $\PC_L$ is upper-bounded by $(L+1)^4$ as shown in \cite{Vidyasagar}.
The \textit{type class}, $T_L$, of a given type, $p\in\PC_L$, is then defined as the set of all length-$L$ sequences whose types are equal to $p$:
\begin{equation}
    T_L(p)=\left\{\xB\in\XC^L:\ \px =p\right\}
\end{equation}
\par
To facilitate analysis, we use a \textit{cyclical} Markov model, where three artificial transitions are added from the end of the sequence to the beginning. 
This model ensures that $\px$ is \textit{consistent}. 
More precisely, for $\aB\in\XC^3$,
\begin{equation}
    \label{consistent}\sum_{b\in\XC}\px(\aB b)=\sum_{b\in\XC}\px(b\aB).
\end{equation}
Note that this implies $\px\in\widetilde P$ as defined in Section \ref{sec:problem_statement}. Furthermore, the third-order and conditional empirical distributions can be derived from $\px$ as follows, for any $b\in\XC$,
\begin{align}
    \label{third_order} \px(\aB )&=\sum_{b\in\XC}\px(\aB b)
\intertext{and}
    \label{conditional}\px(b\vert \aB)&=\frac{\px(\aB b)}{\px(\aB)}
\end{align}
We now use some Large Deviations theory to make an argument about the probability of error in the hypothesis test.
\par 
\subsubsection{Large Deviations Principle}
Vidyasagar \cite{Vidyasagar} provides an extensive analysis of large deviations theory for Markov processes.
Theorems \ref{thm:likelihood} and \ref{thm:sanovs}, shown below, utilize this analysis along with \cite[Lemma~1]{Gutman}, which allows us to make an argument about the probability of error for the subsequent hypothesis test. 
For the proofs of Theorems \ref{thm:likelihood} and \ref{thm:sanovs}, the reader is referred to \cite[Theorem~7]{Vidyasagar} and \cite[Chapter~11]{Cover}. \par
The results in \cite{Vidyasagar} show that a Markov process $\XB~=~(X_1,\dots,X_L)$ with type $p$ and generated by $q$ satisfies the large deviations property with rate function 
\begin{equation}
    I(p)=\DivCond pq
\end{equation}
Here, $D_c$ is the conditional relative entropy defined as the KL divergences averaged over $p$:
\begin{align*}
    \DivCond pq= & \sum_{\aB\in\XC^3} p(\aB)\sum_{b\in\XC}p(b\vert\aB)\log\left({\frac{p(b\vert\aB)}{q(b\vert\aB)}}\right)\\
    =& \sum_{\aB\in\XC^3,b\in\XC} p(\aB b)\log\frac{ p(\aB b)}{q(\aB b)} \\ 
    & - \sum_{\aB\in\XC^3} p(\aB)\log\frac{ p(\aB)}{q(\aB)} \\
    =& D^{(4)}(p\|q) - D^{(3)}(p\|q)
\end{align*}
i.e. the divergence between the fourth-order distributions minus the divergence between the third-order distributions.
Similarly, the ``Markov conditional entropy" can be written as
\begin{align*}
    \EntCond p= & \sum_{\aB\in\XC^3}p(\aB)\sum_{b\in\XC}p(b\vert\aB)\log p(b\vert\aB)\\
    =& H^{(4)}(p)-H^{(3)}(p)
\end{align*}
We use $D_c$ and $H_c$ so as to distinguish between the normal divergence and entropy.
\begin{theorem}\label{thm:likelihood} The probability of $\xB$ under $q$ depends only on its type $\px$ and is given by 
\begin{equation}
    q^{(L)}(\xB)=2^{-L\left[\DivCond{\px}{q} + \EntCond{\px}\right]+\log\alpha}
\end{equation}
where $\alpha=q(x_1x_2x_3)$, i.e. the probability of the initial state of $\xB$.
\end{theorem}
\begin{theorem} [Sanov's Theorem for Markov Processes]\label{thm:sanovs}
Let $\XB=(X_1,X_2,\dots,X_L)$ be a Markov process $q$, and let $\FC\subseteq\widetilde{\PC}$.
The probability that the empirical distribution of $\XB$ is contained in $\FC$, denoted $q^{(L)}(\FC)$, is upper-bounded as
\begin{equation}
    q^{(L)}(\FC)\leq\left|\PC_L\right| 2^{-L \DivCond{p^*}{q}+\log\alpha}
\end{equation}
where $p^*$ is the information projection of $q$ onto $\FC$:
\begin{equation}\label{p_star_sanovs}
    p^*=\arginf_{p\in\FC} \DivCond pq
\end{equation} 
If, in addition, the closure of $\FC$ is equal to the closure of its interior ($\bar{\FC}=\overline{\FC^o}$), then 
\begin{equation} \label{eq:sanov_closure_of_interior}
    \lim_{L\rightarrow\infty}\frac1L \log q^{(L)}(\FC)=-\DivCond{p^*}{q}=-I(p^*)
\end{equation}
\end{theorem}
\subsubsection{Hypothesis Test}
In the binary hypothesis test, there are two candidate models for $q$: $p_1$ and $p_2$, where $p_1\neq p_2$.
We decide between the two hypotheses:
\begin{itemize}
    \item $\text H_1:\ q=p_1$
    \item $\text H_2:\ q=p_2$
\end{itemize}
\par
Let $\PC_1$ and $\PC_2$ be the decision regions for H$_1$ and H$_2$, respectively. 
The sets $\PC_1$ and $\PC_2$ form a partition of $\widetilde{\PC}$ ($\PC_1\cup\PC_2=\widetilde{\PC}$).
As a result, given any $\xB\in\XC^L$, $\delta_N(\xB)$ decides H$_1$ if $\px\in\PC_1$ and H$_2$ if $\px\in\PC_2$. 
The Bayesian probability of error, $P_e$, for the binary hypothesis test with priors $\pi_1$ and $\pi_2$ is given by
\begin{align} 
    P_e = \pi_1 p_1^{(L)}(\PC_2)+\pi_2 p_2^{(L)}(\PC_1)
\end{align}
To minimize the error, the decision rule\footnote{The decision rule $\delta_L$ uses overloaded notation with the decision rule for the main problem.} uses a Neyman-Pearson test
\begin{align}
    \delta_N(\px) =
        \begin{cases}
            \text H_1 & \text{if } \LC(\xB)\geq\frac{\pi_2}{\pi_1} \\
            \text H_2 & \text{if } \LC(\xB)<\frac{\pi_2}{\pi_1}
        \end{cases}
\end{align}
where the likelihood ratio, $\LC$, is defined as:
\begin{equation}\label{lrt}
    \LC(\xB) =
    \frac
        {\text{Pr}(\xB\vert\text H_1\text{ true})}
        {\text{Pr}(\xB\vert\text H_2\text{ true})}
    = \frac
        {p_1^{(L)}(\xB)}
        {p_2^{(L)}(\xB)}
\end{equation}
Using Theorem \ref{thm:likelihood}, the normalized log-likelihood ratio is
\begin{align*}\label{loglrt}
    \frac1L\log\LC(\xB) =& -\left[\DivCond{\px}{p_1} + \EntCond{\px}\right] + \frac{\log\alpha_1}L \\
    & +\left[\DivCond{\px}{p_2}+\EntCond{\px}\right]-\frac{\log\alpha_2}L \\
    =& \DivCond{\px}{p_2} - \DivCond{\px}{p_1}+\frac1L\log\frac{\alpha_1}{\alpha_2}
\end{align*}
\noindent
Again, $\alpha_1$ and $\alpha_2$ represent the probabilities of the initial states of $\xB$ under $p_1$ and $p_2$, respectively.
Notice that as $L~\to~\infty$, the optimal decision rule simply chooses $\argmin\limits_{k\in\{1,2\}}~\DivCond{p}{p_k}$ because the effect of the priors washes out with $L$, along with the probability of the initial states. 
We will show that, by using the decision regions, $\PC_1$ and $\PC_2$, given by
\begin{align}
    \PC_1&=\{p\in\widetilde{\PC}:\DivCond{p}{p_2} - \DivCond{p}{p_1}\geq0\}\\
    \PC_2&=\{p\in\widetilde{\PC}:\DivCond{p}{p_2} - \DivCond{p}{p_1}<0\}
\end{align}
the optimal error exponent is achieved in the limit. 
First we will prove Lemmas \ref{lemma:convexity} and \ref{lemma:closure_of_interior}, which allow for the use of (\ref{eq:sanov_closure_of_interior}) in Theorem \ref{thm:sanovs}.
\begin{lemma}\label{lemma:convexity}
$\PC_1$ and $\PC_2$ are convex.
\begin{proof}
Let $p_a,p_b\in\PC_1$ and let $$p_{ab}=\lambda p_a + (1-\lambda)p_b,\quad\lambda\in(0,1)$$ be a convex combination of $p_a$ and $p_b$. 
Then
\begin{align*}
    &\quad\ \DivCond{p_{ab}}{ p_2}-\DivCond{p_{ab}}{ p_1} \\
    &= \sum_{\aB\in\XC^3,b\in\XC}p_{ab}(\aB b)\log{\frac{p_{ab}(b\vert\aB)}{ p_2(b\vert\aB)}} \\ & \quad - \sum_{\aB\in\XC^3,b\in\XC}p_{ab}(\aB b)\log{\frac{p_{ab}(b\vert\aB)}{ p_1(b\vert\aB)}} \\
    &= \sum_{\aB\in\XC^3,b\in\XC}p_{ab}(\aB b)\log{\frac{ p_1(b\vert\aB)}{ p_2(b\vert\aB)}} \\
    &= \lambda \left[\DivCond{p_a}{ p_2}-\DivCond{p_a}{ p_1}\right] \\
    &\quad + (1-\lambda) \left[\DivCond{p_b}{ p_2}-\DivCond{p_b}{ p_1}\right] \geq0 \notag
\end{align*}
so $p_{ab}\in\PC_1$ and therefore $\PC_1$ is a convex set. A similar argument can be made for the set $\PC_2$. 
Note that since $\PC_1$ and $\PC_2$ are both convex, this implies that the boundary linearly divides the set of stationary fourth-order distributions, $\widetilde{\PC}$.
\end{proof}
\end{lemma}
\begin{lemma}\label{lemma:closure_of_interior}
\begin{equation}\label{closure_of_interior}
    \overline{\PC_1} = \overline{\PC_1^o} \quad \text{and} \quad \overline{\PC_2} = \overline{\PC_2^o}
\end{equation}
\begin{proof}
The boundary between $\PC_1$ and $\PC_2$ consists of the set of distributions $p\in\widetilde{\PC}$ for which $\DivCond{p}{ p_2}-\DivCond{p}{ p_1}=0$. 
We see that $ p_1$ does not lie on this boundary because
\begin{align}
  \DivCond{ p_1}{ p_2}-\DivCond{ p_1}{ p_1}=\DivCond{ p_1}{ p_2}> 0
\end{align}
by the non-negativity of the KL-divergence. 
Furthermore, $ p_1$ cannot lie on any other boundary of $\PC_1$ because all of the elements of $ p_1$ are nonzero. Thus $ p_1$ is an interior point of $\PC_1$ as it does not lie on any of the boundaries.
\par
Finally, we need to show that convexity and a non-empty interior imply (\ref{closure_of_interior}). 
Take a point $p\in\overline{\PC_1}$. Then either $p\in\PC_1^o$ or $p\in\partial\PC_1$, the boundary of $\PC_1$.
If $p\in\PC_1^o$, then $p\in\overline{\PC_1^o}$, trivially.
If $p\in\partial\PC_1$, we must prove that $p$ is a limit point of $\PC_1^o$. 
Since $ p_1$ is an interior point of $\PC_1$, then there exists an open ball $U_1$ centered at $ p_1$ which is completely contained in $\PC_1$.
We define $V_1$ as the set of distributions that result from a convex combination of $p$ and $U_1$
\begin{equation}
    V_1 = \{\alpha U_1+(1-\alpha)p:0<\alpha\leq1\}
\end{equation}
using Minkowski addition. The set $V_1$ clearly has non-zero volume (by Lebesgue measure) and all of its points are interior points of $\PC_1$ due to Lemma \ref{lemma:convexity}. 
Therefore there exists a sequence of interior points $\{p_t\},p_t\in V_1$ such that $p_t\rightarrow p$. Thus  $p\in\overline{\PC_1^o}$. 
\par
A similar argument can be made for $\PC_2$. Hence, the proof of Lemma \ref{lemma:closure_of_interior} is complete.
\end{proof}
\end{lemma} 
Now, by Theorem \ref{thm:sanovs} and Lemmas \ref{lemma:convexity} and \ref{lemma:closure_of_interior}, the error exponents are
\begin{align}
    \lim_{L\rightarrow\infty}\frac1L\log p_1^{(L)}(\PC_2)=& -\DivCond{p_1^*}{p_1} \\
    \lim_{L\rightarrow\infty}\frac1L\log p_2^{(L)}(\PC_1)=& -\DivCond{p_2^*}{p_2}.
\end{align}
Distribution $p_1^*$ is found by minimizing $\DivCond{p_1^*}{p_1}$, subject to the decision boundary constraint,
\begin{equation} \label{eq:constraint}
    \DivCond{p_1^*}{ p_1} - \DivCond{p_1^*}{ p_2} \geq 0,
\end{equation}
the consistency constraints for all $a\in\XC^3$,
\begin{equation}
    \sum_{b\in\XC}p_1^*(\aB b)=\sum_{b\in\XC}p_1^*(b\aB),
\end{equation}
and the sum-to-one constraint,
\begin{equation}
    \sum_{\cB\in\XC^4}p_1^*(\cB)=1
\end{equation}
This will yield the distribution $p_1^*\in\PC_2$ that is closest to $p_1$.
Moreover, we claim that $p^*$ must lie on the boundary, i.e. (\ref{eq:constraint}) holds with equality.
This can be proven by contradiction: suppose $p'$ is the optimal solution to the minimization problem and suppose $$D_c(p'\|p_1)-D_c(p'\|p_2)>0.$$
For $0~\leq~\lambda~\leq~1$, let $p_\lambda~=~\lambda~p_1~+~(1-\lambda)~p'$ be a convex combination of $p'$ and $p_1$. 
We know from Lemma \ref{lemma:convexity} that $p_\lambda~\in~\widetilde{\PC}$ for any value of $\lambda$ and furthermore, there exists a $\lambda=\lambda^*$ such that $$D_c(p_{\lambda^*}\|p_1)-D_c(p_{\lambda^*}\|p_2)=0$$ since the boundary linearly divides $\widetilde{\PC}$.
Now, to show by contradiction that $$D_c(p_{\lambda^*}\|p_1)<D_c(p'\|p_1),$$ we will show that conditional relative entropy is convex in its first argument. For some distribution $q\in\widetilde{\PC}$,
\begin{align*}
    D_c(p_{\lambda}\|q) &= \sum\limits_{\aB b\in\XC^4}p_\lambda(\aB b)\log\frac{p_\lambda(b|\aB)}{q(b|\aB)} \\
    &= \lambda\sum\limits_{\aB b\in\XC^4}p_1(\aB b)\log\frac{p_\lambda(b|\aB)}{q(b|\aB)} \\ &\quad+ (1-\lambda)\sum\limits_{\aB b\in\XC^4}p'(\aB b)\log\frac{p_\lambda(b|\aB)}{q(b|\aB)} \\
    &= \lambda\sum\limits_{\aB b\in\XC^4}p_1(\aB b)\LR({\log\frac{p_1(b|\aB)}{q(b|\aB)}-\log\frac{p_1(b|\aB)}{p_\lambda(b|\aB)}}) \\ 
    &\quad + (1-\lambda)\sum\limits_{\aB b\in\XC^4}p'(\aB b)\LR({\log\frac{p'(b|\aB)}{q(b|\aB)}-\log\frac{p'(b|\aB)}{p_\lambda(b|\aB)}}) \\
    &= \lambda D_c(p_1\|q)+(1-\lambda)D_c(p'\|q) \\ &\quad -\lambda D_c(p_1\|p_\lambda)-(1-\lambda)D_c(p'\|p_\lambda) \\
    &<\lambda D_c(p_1\|q)+(1-\lambda)D_c(p'\|q)
\end{align*}
where the last step follows from the non-negativity of conditional relative entropy. Finally, setting $q=p_1$, we have $$D_c(p_\lambda\|p_1)<\lambda D_c(p_1\|p_1)+(1-\lambda)D_c(p'\|p_1)<D_c(p'\|p_1).$$ Therefore, $p_{\lambda^*}$ must be a better solution, which is a contradiction. Clearly $p_1^*=p_2^*$ since they both lie on the boundary where the minimands, $D_c(\cdot\|p_1)$ and $D_c(\cdot\|p_2)$, are equal.


\end{document}